\documentclass{revtex4}
\usepackage{graphicx}
\usepackage{dcolumn} 
\usepackage{bm}
\usepackage{amsmath}
\usepackage{multirow}
\begin{document}
\title{What can we learn from recent $2\nu\beta\beta$ experiments?}
\author{Dong-Liang Fang} 
\address{ Institute of Modern Physics, Chinese Academy of sciences, Lanzhou, 730000, China}
\address{ University of Chinese Academy of Sciences, Beijing, 100049,China}
\begin{abstract}
With recent measurements of the two neutrino double beta decay high precision electron spectra, combining with charge exchange or $\beta$-decay experimental data, we give severe constraints over the current nuclear many body calculations. Our calculation shows that Quasi-particle Random Phase Approximation (QRPA) approach can well reproduce the measured spectra for the two open shell nuclei, $^{82}$Se and $^{100}$Mo. For the closed shell nucleus $^{136}$Xe, QRPA can also reproduce the spectra with proper treatments. We also find that considering the high-lying state reduction, Nuclear Shell Model can also well reproduce the spectra as well as Gamow-Teller transition strength under a unique quenched Axial-vector coupling constant $g_A$. For $^{136}$Xe, we find that the flip of the sign for the decay strength will lead the spectra to go beyond the so-called High-lying state dominance hypothesis. 
These results call for future high precision measurement of charge-exchange reaction.  
\end{abstract}
\maketitle
\section{introduction}
For decades, more and more double beta decay experiments have been proposed and constructed. They aim to find the unique mode of a second order weak decay -- double beta decay ( hereafter $\beta\beta$-decay ), the so-called neutrinoless double beta decay ($0\nu\beta\beta$). The discovery of this decay mode could lead us to the new era of new physics beyond Standard Model. As a byproduct, experimental data of the observed mode of $\beta\beta$-decay, namely the two neutrino double beta decay ($2\nu\beta\beta$), has also been accumulated. With the help of these data, we could achieve better understanding of the nuclear structure aspect for the decay and improve the predictive power of the theory. An example is the combined study to determine the $g_A$ quenching behavior from the electron spectra\cite{Simkovic:2018rdz} from Kamland-Zen experiment\cite{KamLAND-Zen:2019imh}. These data can also be used to probe possible new physics behind $2\nu\beta\beta$, such as the Lorentz violation in the decay\cite{Ghinescu:2022vwd} or the implications of right-hand weak current\cite{Deppisch:2020mxv}.

High precision electron spectra of $2\nu\beta\beta$ for at least three nuclei: $^{82}$Se\cite{Arnold:2018tmo,Azzolini:2019yib}, $^{100}$Mo\cite{NEMO-3:2019gwo} and $^{136}$Xe\cite{KamLAND-Zen:2019imh} have been measured. And among the three, a quantitative constraint is given for $^{136}$Xe. For $^{82}$Se, the summed spectra show strong favors on the so-called single state dominance(SSD) which indicates that the first intermediate states contribute enough decay strength. Similar results are obtained for $^{100}$Mo with even more significant implications that the first intermediate state may contribute more strength than that needed to reproduce the decay half-lives. However, this behavior of SSD is not universal, results from Kam-Land Zen suggests that for $2\nu\beta\beta$ of $^{136}$Xe, SSD is strongly disfavored and is excluded at the confidence level of 97\%. 
Previously, measurements of half-lives provide the nuclear matrix elements which then served as constraints for fixing the model parameter for various calculations. Could these new measurements be used to further constrain nuclear structure calculations? The answer is definitely yes, and such attempts have been already done by \cite{Simkovic:2018rdz}.

In traditional simulations of $2\nu\beta\beta$ decay half-lives as well as spectra, to separate the lepton and nuclear parts, one uses averaged nuclear excitation energies in the the phase space factor (PSF) calculations and equal electron mass Ansatz for calculating nuclear matrix elements (NME). Under these assumptions, the half-lives come out to be the products of PSF and NME, while the spectra depends on PSF only. The uncertainties for PSF then come from the choice of the average nuclear excitation energy. A convenient assumption is that the decay is dominated by a single state, usually the first states or high-lying GTR states, they are called SSD mentioned above or high-lying states dominance (HSD) respectively. Beyond above calculations, there are also calculations with fully numerical treatment that the nuclear and lepton parts are calculated simultaneously\cite{Moreno:2008dz}, and in this case no artificial average nuclear excitation energy is needed. As a consequence, the spectra are now related to the nuclear structure. However, their calculations for $^{100}$Mo eliminate the existence of SSD and obtain a spectrum much closer to HSD\cite{Moreno:2008dz}. This somehow contradicts recent measurement\cite{NEMO-3:2019gwo}. Therefore, unlike the NME which could always be fixed by adjusting the model parameters, the electron spectra put more stringent constraints on the calculation and could possibly rule out certain calculations.

Different from neutrinoless double beta decay calculations, the intermediate states are important for two neutrino double beta decay. And till present, the methods widely used in $2\nu\beta\beta$ calculations are Quasi-particle Random Phase Approximation (QRPA) and Nuclear Shell Model (NSM). For QRPA, with proper choice of the parameter $g_{pp}$ and $g_A$ we can fit the nuclear matrix elements as indicated above, but there is still questions of simultaneously fulfill the double beta decay and beta decay nuclear matrix elements\cite{Faessler:2007hu}, therefore, whether QRPA could reproduce double beta decay transition process is still doubtable. In this sense, the electron spectra can be used as an alternative to check the correctness of QRPA calculations. For NSM, except for $^{48}$Ca, other nuclei can only be calculated with a spin-orbit partner incomplete model space. Such a model space will not fulfill the non-energy-weighed Ikeda sum rule, therefore one usually needs to introduce further quenching of the transition strength\cite{Brown:2015gsa}. Meanwhile, for such a model space, the GTR states, coming mostly from transitions between spin-orbit partner orbitals, are missing. As a result, severe reduction to the strength of $\beta\beta$-decay via GTR predicted by QRPA is missing too. Could the study of electron spectra provide us information of whether these reductions are actually presented? This will be answered from current study. 
 
This work is arranged as follows: in the first part a numerical formalism is revisited, and then the results and discussions are followed by the conclusion.
\section{Formalism}
Starting from the S-matrix theory, one can obtain the $2\nu\beta\beta$ decay width generally expressed as\cite{Doi:1985dx}:
\begin{eqnarray}
d\lambda^{2\nu}&=&(a^{(0)}+a^{(1)}\cos\theta_{12})\omega_{2\nu}d\epsilon_1d\epsilon_2d\omega_1 d\cos\theta_{12}
\end{eqnarray}
For this expression, we consider only contribution from lepton's $s-$ partial wave, those of other partial waves can be safely neglected since they are suppressed by small lepton momenta. $\epsilon_{1,2}$ and $\omega_{1,2}$ are electron and neutrino energies respectively. $\theta_{12}$ is the angle between the two emitted electrons, the $a^{(1)}$ term describes the angular distribution of the electrons, since it doesn't contribute to the final decay width and the electron energy spectra, we ignore them in subsequent discussion.

The kinematic factor $\omega_{2\nu}$ is:
\begin{eqnarray}
\omega_{2\nu}=\frac{G\cos^4\theta_c}{64\pi^7} \omega_{1}^2\omega_{2}^2 p_1 p_2 \epsilon_1 \epsilon_2
\end{eqnarray}
Here, $G$ is the Fermi constant and $\theta_C$ is the Cabibbo angle. $p_{1(2)}$ are the momenta of the two electrons.

The transition strength $a^{(0)}$ can be expressed as \cite{Doi:1985dx}:
\begin{eqnarray}
a^{(0)}&=& f_{11}^{(0)} (|g_A A_{GT}^{+}-g_V A_{F}^+|^2+\frac{1}{3}|g_A A_{GT}^{-}+3 g_V A_{F}^-|^2) 
\label{str}
\end{eqnarray}
With $F$ for Fermi and $GT$ for Gamow-Teller transitions.

The related NME's are defined as: 
\begin{eqnarray}
A_{I}^{\pm}=\frac{1}{2}\sum_N \langle f | \mathcal{O}^{f}_I|N\rangle \langle N |\mathcal{O}^i_I |0_i^+\rangle(K_N\pm L_N) 
\label{ME}
\end{eqnarray}
With
\begin{eqnarray}
K_N&=&\frac{1}{E_N+(\epsilon_1+\omega_1-\epsilon_2-\omega_2)/2}-\frac{1}{E_N-(\epsilon_1+\omega_1-\epsilon_2-\omega_2)/2} \nonumber \\
L_N&=&K_N(\epsilon_1 \rightleftharpoons \epsilon_2,\omega_1 \rightleftharpoons \omega_2)
\label{KL}
\end{eqnarray} 
the energy denominators. 

$E_N=E^{ex}_N+(2M_m-M_i-M_f)/2$ is the average of the energy differences between the $N$th intermediate states and the even-even initial and final states, with $E_N^{ex}$ is the excitation energy of the Nth state and $M_m$, $M_i$ and $M_f$ are the masses for the intermediate, initial and final nuclei respectively. The nuclear transition  operators are $\sigma$ for GT and $1$ for Fermi matrix elements respectively. It is known, $M_{F}^{\pm}$ vanishes due to isospin symmetry. The expression can then be reduced to:
\begin{eqnarray}
a^{(0)}&=& f_{11}^{(0)} g_A^2 |M_{GT}^{(0)}(\epsilon_1,\epsilon_2,\omega_1,\omega_2)|^2
\end{eqnarray} 
With the lepton energy dependent NME's $M_{GT}^{(0)}=(A_{GT}^{+}-A_{GT}^{-}/3)/2$. 

The electron part of the transition has the form under the so-called no-finite de-Broglie wavelength correction(no-FBWC)\cite{Doi:1985dx,Kotila:2012zza}:
\begin{eqnarray}
f_{11}^{(0)}&=&|f^{-1-1}|^2+|f_{11}|^2 + |f^{-1}{}_1|^2+|f_{1}{}^{-1}|^2 
\end{eqnarray}
With
\begin{eqnarray}
f^{-1-1}&=&g_{-1}(\epsilon_1,R) g_{-1}(\epsilon_2,R) \nonumber \\
f_{11}&=&f_1(\epsilon_1,R) f_1(\epsilon_2,R) \nonumber \\
f^{-1}{}_{1}&=&g_{-1}(\epsilon_1,R) f_{1}(\epsilon_2,R) \nonumber \\
f_{1}{}^{-1}&=&f_1(\epsilon_1,R) g_{-1} (\epsilon_2,R) \nonumber
\end{eqnarray}
Here the electron radial wave-functions (upper component $g_\kappa$'s and lower component $f_\kappa$'s) are obtained by solving the Dirac equations and $R$ here refers to the empirical nuclear radius $R=1.2A^{1/3}$fm. The subscripts of the electron wave functions $\kappa$ are defined in literature\cite{Doi:1985dx} to distinguish different spherical partial waves for electron.

While the lepton part, the electron wave functions, can be calculated with a decent accuracy, the calculation of nuclear part (the NME) is always a tough task, limited precision can be achieved. In current study, we adopt two many-body approaches: QRPA and NSM for this part. 

For NSM calculations, we use the NuShellX@MSU\cite{Brown:2014bhl} code. And by diagonalizing the Hamiltonian we obtain the wave functions for states of the initial, final and intermediate nuclei. We then get the NME [Eq.\eqref{ME}] with the reduced one-body transition amplitude from states A to B and the corresponding single particle matrix element 
\begin{eqnarray}
\langle B || \sigma\tau^+||A\rangle=\sum_{pn}\langle B|| [c_{p}^{\dagger}c_{\tilde{n}}]_{1^+}|| A\rangle \langle p||\sigma||n\rangle
\end{eqnarray}
The first term in the r.h.s of above formula are the major output from NSM calculations, and the second part can be obtained analytically for Spherical Harmonic Oscillator basis.

For the QRPA method, we adopt the version with realistic forces\cite{Tomoda:1987ji}, it is widely used in the nuclear weak decay calculations. The intermediate states can be constructed as proton-neutron excited states on the even-even BCS ground states, therefore in QRPA we have two sets of intermediate states starting from the initial and final excited states. Based on the BCS ground state, the pn-QRPA phonon can be expressed as $|1^+\rangle=Q^\dagger_{1^+_m}|BCS\rangle$ with $Q^\dagger_{m}\equiv\sum_{pn}(X^m_{pn}\alpha_p^\dagger \alpha_n^\dagger-Y^m_{pn}\tilde{\alpha}_n\tilde{\alpha}_p)$. Here $\alpha$'s are the BCS quasi-particle. And the forward and backward amplitudes $X$ and $Y$'s can be obtained from the QRPA equations, the detailed forms of QRPA equations and the residual interactions are presented in \cite{Simkovic:2013qiy}. With QRPA method, the NME of Eq.\eqref{ME} has the specific form as:
\begin{eqnarray}
A^{\pm}_{GT}=\frac{1}{2}\sum_{m_i,m_f} \langle 0_f^+||\sigma\tau^+||m_f \rangle \langle m_f | m_i \rangle \langle m_i ||\sigma\tau^+||0_i^+\rangle (K_{m_i}\pm L_{m_i})
\end{eqnarray}  
The detailed expression for one-body transition amplitude is presented in for example \cite{Moreno:2008dz}.

Since $|m_i\rangle$ and $|m_f\rangle$ are the same set of states expressed on two different Hilbert basis. We could choose one of which to be the one corresponding to the natural one. In this case, one usually choose $|m_i\rangle$ to be the excitation states corresponding to the actual states. Then, we can express:
\begin{eqnarray}
A^{\pm}_{GT}=\frac{1}{2}\sum_{m_i} \langle 0_f^+||\sigma\tau^+|| m_i \rangle \langle m_i ||\sigma\tau^+||0_i^+\rangle (K_{m_i}\pm L_{m_i})
\end{eqnarray}  
Where for QRPA calculations $\langle 0_f^+||\sigma\tau^+|| m_i \rangle=\sum_{m_f} \langle 0_f^+||\sigma\tau^+||m_f \rangle \langle m_f | m_i \rangle$.
And the $E_N$ terms in K and L and now expressed as functions of $\omega_i$, we define $E^{ex}_N=\omega^{1^+}_{m_i}-\omega^{1^+}_0+E_{1+}^{exp.}$ . 



With these calculated NMEs and electron wave functions, we can define following final decay rates \cite{Doi:1985dx}:
\begin{eqnarray}
F^{(0)} & = & 2\int_{m_e}^{Q+m_e} \int_{m_e}^{Q+m_e-\epsilon_1} \int_{0}^{Q-\epsilon_1-\epsilon_2} a^{(0)}\omega_{2\nu}d\epsilon_1d\epsilon_2d\omega_1 
\label{numexp}
\end{eqnarray}

And normalized differential decay rates or spectra of single and summed electron can be expressed as $dN/d\epsilon_1=dF^{(0)}/d\epsilon_1/F^{(0)}$ and $dN/d(\epsilon_1+\epsilon_2)=dF^{(0)} /d(\epsilon_1+\epsilon_2)/F^{(0)}$ respectively \cite{Doi:1985dx,Kotila:2012zza}.


Compared to current full numerical treatment, previous calculations \cite{Doi:1985dx,Kotila:2012zza,Simkovic:2018rdz} of electron spectra adopt certain approximations for $K_a$ and $L_a$. In \cite{Doi:1985dx,Kotila:2012zza}, by introducing an average excitation energy $\tilde{A}\equiv \langle E_N \rangle $:
\begin{eqnarray}
 K_N &\approx& \frac{\tilde{A}}{E_N}(\frac{1}{\tilde{A} + (\epsilon_1+\omega_1-\epsilon_2-\omega_2)/2} +\frac{1}{\tilde{A} - (\epsilon_1+\omega_1-\epsilon_2-\omega_2)/2})=  \frac{1}{E_N} \tilde{A} \langle K_N \rangle \nonumber \\
 L_N &\approx& \frac{1}{E_N-(M_I+M_F)/2} \tilde{A} \langle K_N(\epsilon_1 \rightleftharpoons \epsilon_2,\omega_1 \rightleftharpoons \omega_2) \rangle = \frac{1}{E_N} \tilde{A} \langle L_N \rangle
\end{eqnarray}
The nuclear and lepton part can then be well separated:
\begin{eqnarray}
a^{(0)}\approx \frac{1}{4} f_{11}^{(0)} |M^{2\nu}_{GT}|^2 \tilde{A}^2 [(\langle K_N \rangle +\langle L_N \rangle)^2+\frac{1}{3}(\langle K_N \rangle -\langle L_N \rangle)^2]
\label{rapp} 
\end{eqnarray}
Here $M_{GT}^{2\nu}=\sum_m [\langle 0^+_f ||\sigma \tau^+ ||m\rangle \langle m || \sigma \tau^+ || 0_i^+ \rangle)/E_m]$ is the nuclear matrix elements for double decay adopted in various literature\cite{Doi:1985dx,Kotila:2012zza,Barabash:2015eza}, and the decay half-life is given as $(T_{1/2}^{2\nu})^{-1}=g_A^4G^{2\nu} |M_{GT}^{2\nu}|^2$. Such an approximation could give a very precise description for decay half-lives, but relies on a hand-insert $\tilde{A}$ description for the electron spectra. This then brings along two kind of choices of the $\tilde{A}$: the Single-State-Dominance (SSD) and the High-lying State Dominance (HSD) hypotheses. The former assumes that the decay strength is brought by the first $1^+$ intermediate states predominantly, while the latter suggests these strength comes from the transition through the high-lying states mostly near the giant resonance. However, this approximation is hard to deal with the cases where several states compete with each other as we shall see.

By using an Taylor expansion over the lepton energies, \cite{Simkovic:2018rdz} goes beyond above approximation, $K_N$ and $L_N$ for the Nth excited state are now expressed as:
\begin{eqnarray}
K(L)_N =  \frac{2}{E_N}(1+\frac{\epsilon^2_{K(L)}}{E_N^2}+...)
\end{eqnarray}
With $\epsilon_K=(\epsilon_1+\omega_1-\epsilon_2-\omega_2)/2$ and $\epsilon_L=(\epsilon_1+\omega_2-\epsilon_2-\omega_1)/2$.

Substitute $K$ and $L$ into the expression of eq.\eqref{str}, one obtain the more precise expression \cite{Simkovic:2018rdz}. By integrating over the lepton momenta and keeping the contributions up to sub-leading order, we have:
\begin{eqnarray}
T^{2\nu}_{1/2} = g_A^4 |M^{2\nu}_{GT}|^2 [G_{0}^{2\nu}+\xi_{31}^{2\nu}G^{2\nu}_2+...]
\label{TEM}
\end{eqnarray}
Here $\xi_{31}^{2\nu}\equiv M^{2\nu}_{GT-3}/(m_eM^{2\nu}_{GT})$, with $M^{2\nu}_{GT-3}\equiv 4m_e^3\sum_{m} [\langle 0^+_f ||\sigma \tau^+ ||m\rangle \langle m || \sigma \tau^+ || 0_i^+ \rangle)/E_m^3]$. The expression of $G^{2\nu}_2$ is also given in\cite{Simkovic:2018rdz}. This approach doesn't need an artificial $\tilde{A}$ , but needs the inputs from nuclear structure calculations. In some case, it could give us very precise description of the spectra. In one word, this method gives a parametrized description for the electron spectra beyond the simplest approximations.

Above two expressions are all approximations of the exact expression [eq.\eqref{numexp}]. We will discuss in subsequent sections about the reliability of these expressions.  

\section{Results and discussion}
As we have shown above, the numerical calculations involve two parts, the lepton part and the nuclear part. The electron wave functions are solutions of Dirac equations under nuclear static electric potentials. In current study, the Dirac equations are solved numerically with the package $\mathrm{RADIAL}$\cite{SALVAT1995151}. The input of the subroutines -- the static Coulomb potential is obtained by assuming an uniformly distributed nuclear charge with charge radius being the empirical nuclear radius.

For the nuclear part, as we have stated above, two approaches, NSM and QRPA, are used respectively. In this study we investigate the spectra of three nuclei: $^{82}$Se, $^{100}$Mo and $^{136}$Xe. And NSM is applicable for $^{82}$Se and $^{136}$Xe, for the former nucleus, we adopt the $\mathbf{jj44}$ model space which consists of 4 levels: $0f_{5/2}$, $1p_{3/2}$, $1p_{1/2}$ and $0g_{9/2}$ for both neutrons and protons, while for the latter nucleus, we use the $\mathbf{jj55}$ model space: $0g_{7/2}$, $1d_{5/2}$, $2s_{1/2}$, $1d_{3/2}$, $0h_{11/2}$, for both neutrons and protons. These model spaces are severely truncated, especially the spin-orbit partner of several important orbitals are missing, therefore the Ikeda sumrule is severely violated. For these model spaces, a drastic quenching on the transition strength may be needed. For $2\nu\beta\beta$, this is usually done with the treatment in \cite{Brown:2015gsa} by comparing the calculated NMEs with the experimental ones. For $^{82}$Se, we use two interactions, namely $\mathbf{jun45}$\cite{Honma:2009zz} and $\mathbf{jj44b}$\cite{Lisetskiy:2004xp}. While for $^{136}$Xe, $\mathbf{jj55a}$\cite{Brown:2004xk} is adopted.

For the QRPA part, we use the nuclear wave functions solved from Schr\"odinger equations with Coulomb corrected Woods-Saxon potential. The quasi-particles are obtained from BCS theory with residue interactions of realistic CD-Bonn force. The same interaction is used to obtain the pn-QRPA phonons. 

For the experimental data, for $^{82}$Se\cite{Arnold:2018tmo,Azzolini:2019yib} and $^{100}$Mo\cite{NEMO-3:2019gwo}, we have only qualitative conclusions which strongly prefer the SSD especially for the latter. However, for $^{136}$Xe, quantitative analysis is available, which rules out the SSD at the C. L. of 97\%. These data can now be used to constrain the many-body calculations in many aspects, such as as those suggested in \cite{Simkovic:2013qiy} to measure the quenching of $g_A$. And also in this work, we want to show, these data can also be used as verifications for nuclear many-body approaches, and benchmark various many-body models and revel more details about the decay such as the existence of cancellation to decay strength from high-lying states.

The numerical treatment described in above section has been used in \cite{Moreno:2008dz} and their QRPA calculations with Skyrme interactions for $^{100}$Mo, $^{116}$Cd and $^{130}$Te, suggest that the calculated spectra is close to HSD, this somehow contradicts recent measurements for $^{100}$Mo\cite{NEMO-3:2019gwo}. Meanwhile, the Shell Model calculations for these nuclei show different trends and the strength sums up as energy increases, this is observed especially for $^{82}$Se, and the $2\nu\beta\beta$ NME seems to converge at energies around $6-7$ MeV. However, the running sum for $^{48}$Ca is quite different\cite{Horoi:2007xe}, an obvious cancellation at high energy is observed as commonly observed in QRPA calculations\cite{Fang:2009ic,Lv:2021oty}. This is due to the fact that for $^{48}$Ca case, a spin-orbit partner complete model space is used and the cancellation is most probably coming from the transitions between spin-orbit partner orbitals. Next we will give detailed studies for each nucleus.


\subsection{$^{82}$Se}

\begin{figure*}
\includegraphics[scale=0.5]{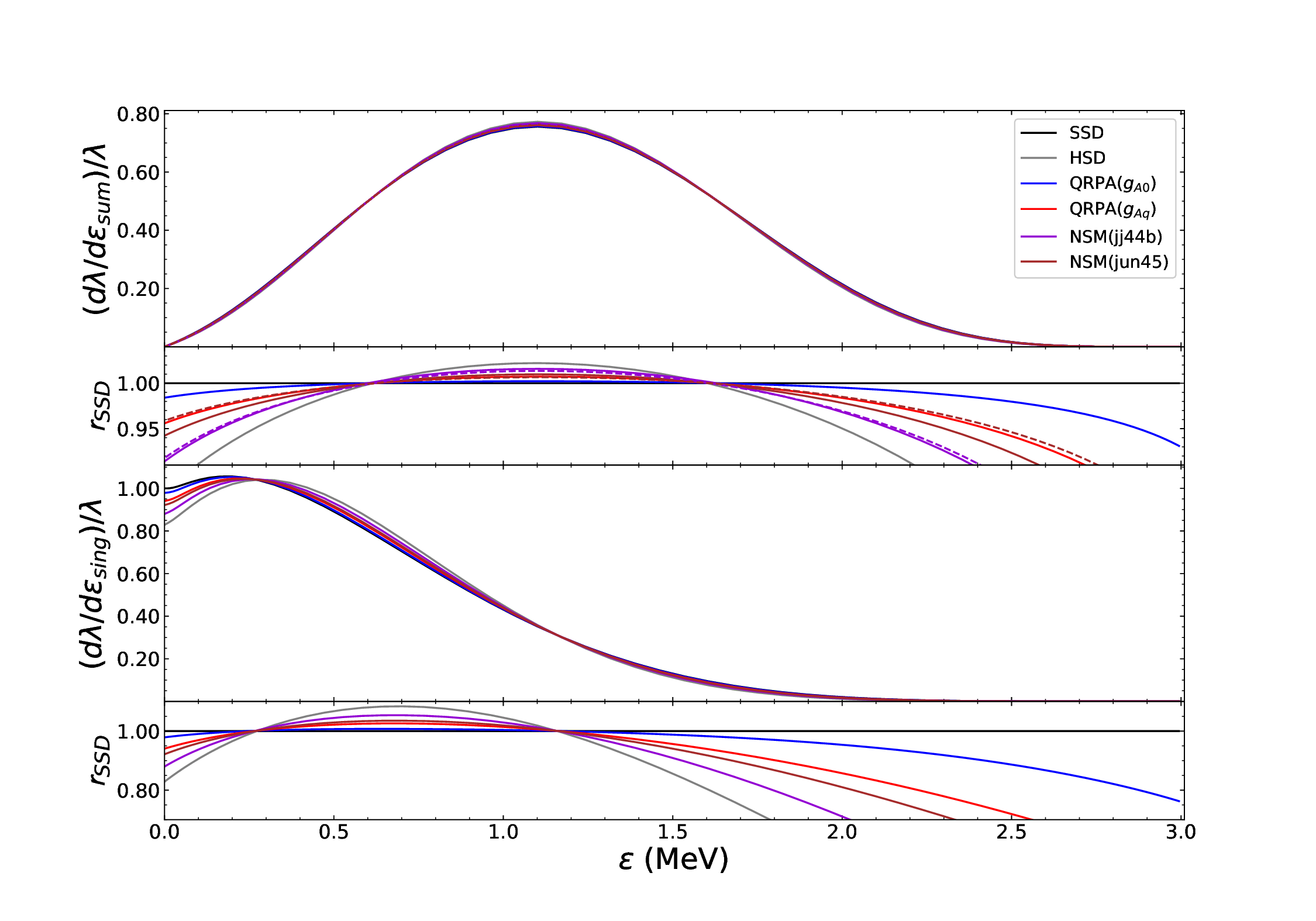}
\caption{(Color online) The electron spectra from NSM and QRPA calculations for $^{82}$Se. NSM calculations are with two Hamiltonians as indicated in the text, while for QRPA, we consider two cases: $g_{A0}$ without $g_A$ quenching and $g_{Aq}$ with $g_A=0.75g_{A0}$. The dashed lines for NSM calculations are for case considering only first few states accumulating enough strength (see the text and also fig.\ref{runsum}). The first and third panels are the summed and single electron spectra respectively, while the second and fourth panels are ratios of each case over the spectra obtained from SSD hypothesis.}
\label{spectraSe}
\end{figure*}

\begin{figure*}
\includegraphics[scale=0.5]{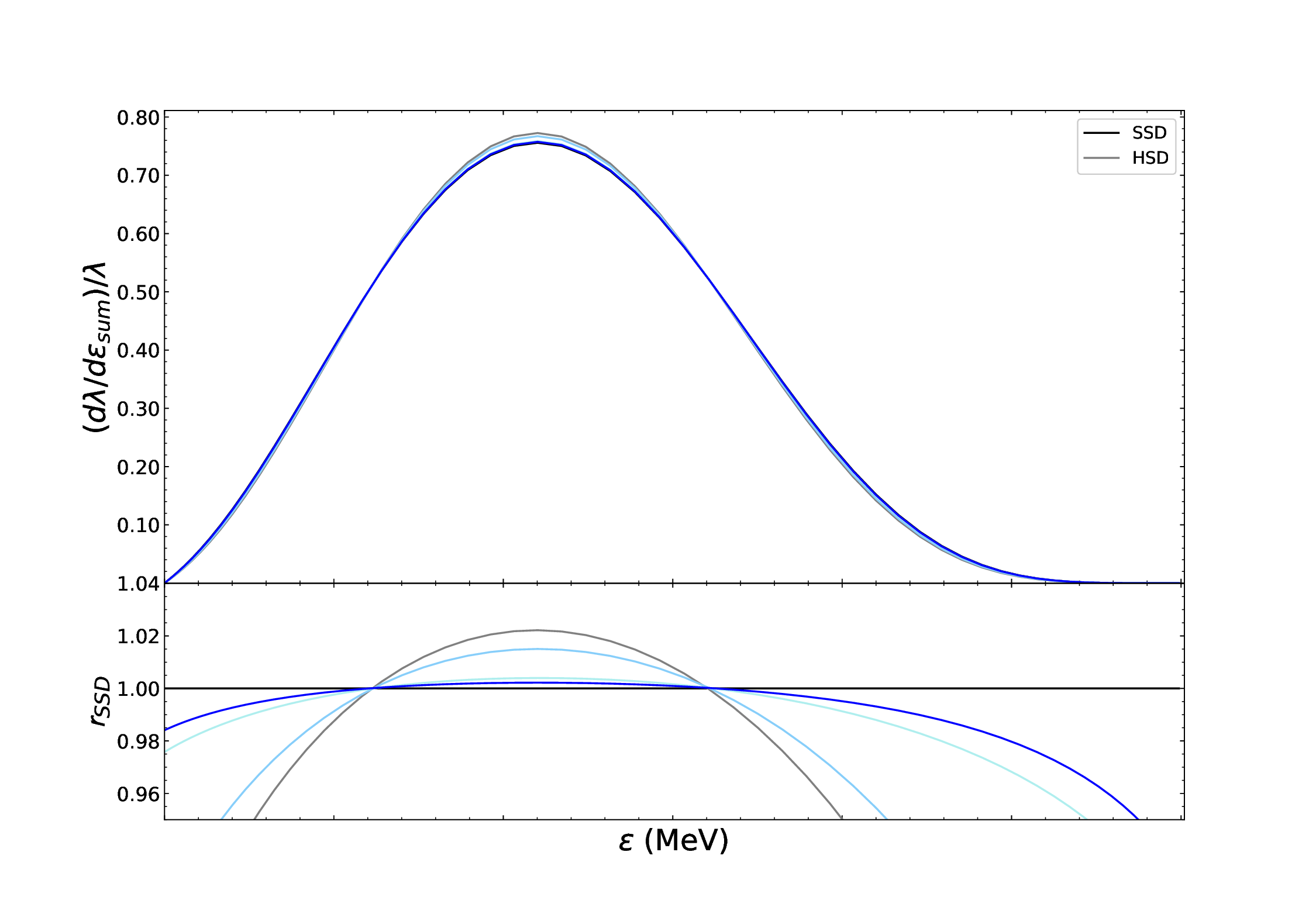}
\caption{(Color online) Illustration of how different states contribute to the summed spectra from QRPA calculations, the colors are in correspondence with fig.\ref{runsum}, the blue lines are with all intermediate states taken into account, while the light blue refers to the exclusion of high-lying states lead to the cancellation, meanwhile the pale blue curve are with low-lying states which accumulate enough strength to reproduce the half-life (see also fig.\ref{runsum}). Here, the upper panel are the spectra and lower panel with $r_{SSD}$ label refers to the ratio of different normalized spectra divided by the normalized SSD spectrum.}
\label{Seind}
\end{figure*}

\begin{figure*}
\includegraphics[scale=0.5]{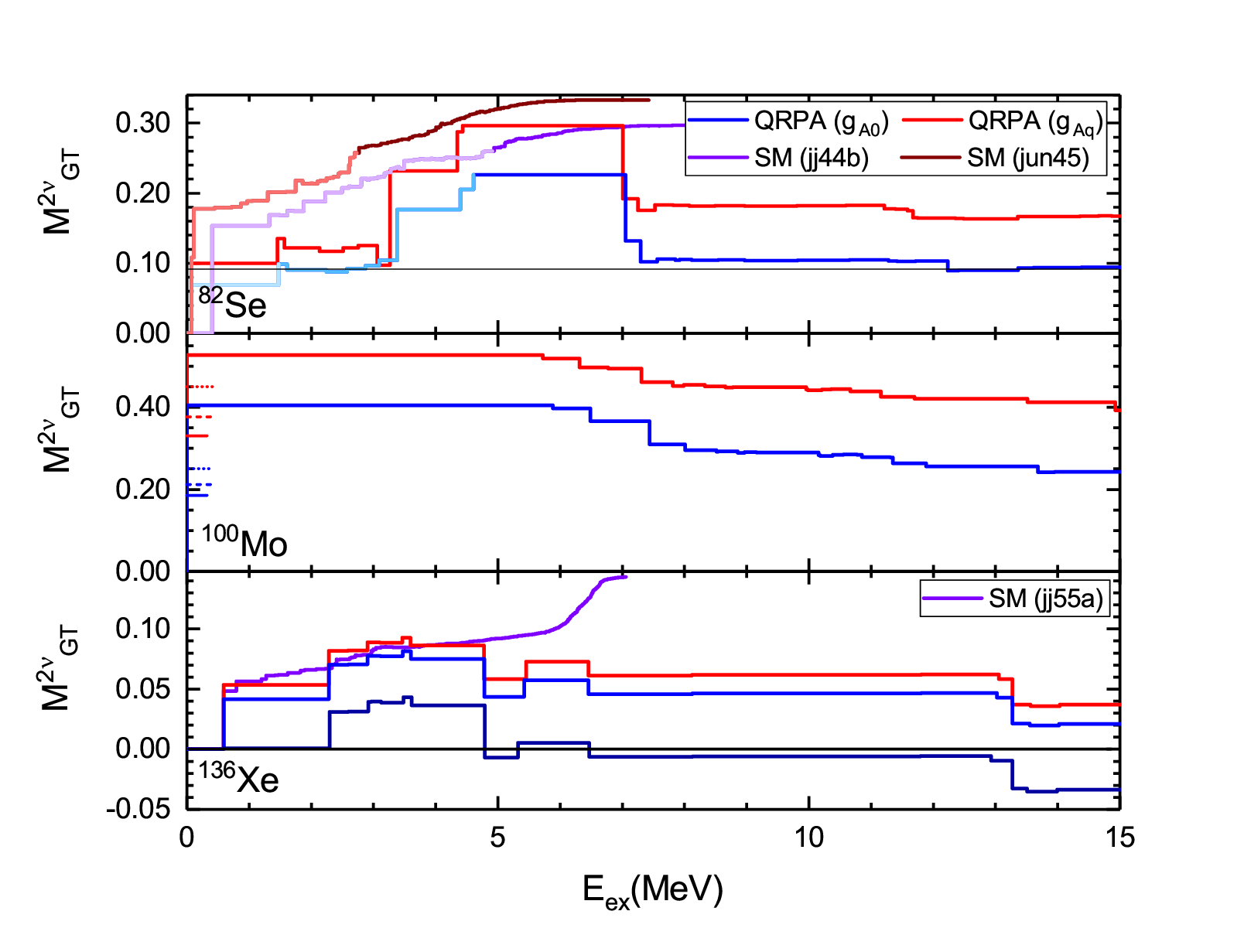}
\caption{(Color online) Running sum of $M^{2\nu}_{GT}$ for $^{82}$Se, $^{100}$Mo and $^{136}$Xe. For $^{82}$Se, results for QRPA and NSM are presented and different colors for QRPA without $g_A$ quenching is explained in the text, so that for NSM. For $^{100}$Mo, QRPA results are presented and the short horizontal lines refers to the experimental results from \cite{Moreno:2008dz,Semenov:2017dgj} by assuming SSD. For $^{136}$Xe, we have an extra dark blue line for QRPA which is explained in the text.}
\label{runsum}
\end{figure*}



For this nucleus, NEMO-3 measurement\cite{Arnold:2018tmo} favors the SSD hypothesis with $\chi^2/$ndf=12.34/16 compared to HSD with $\chi^2$/ndf=35.32/16 for the single electron spectrum. Here, both SSD and HSD are the languages from the naive description of spectra from the roughest approximation (eq.\eqref{rapp}). Our results for this nucleus are presented in fig.\ref{spectraSe}, as a first glance, it suggests result without $g_A$ quenching (Blue line) is much closer to SSD than the quenched case (red line) for QRPA, and in general, QRPA favors SSD rather than HSD. Meanwhile, for NSM calculations (bold lines), $jj44b$ (Purple line) strongly favors HSD and $jun45$ (Brown line) lies in between that of $jj44b$ and QRPA results.

Before proceeding to the detailed discussion of the spectra for this nucleus, we first use this nucleus as an example to show how the different intermediate states shape the electron spectra, to learn if it is possible to understand the spectra with the running sum from the intermediate states of double beta decay. To understand how different states contribute to the final spectra, we plot in fig.\ref{Seind} comparison of spectra including only specific states (In this work, we categorize the states by low-, medium- and high-lying states). 

To explain the choices of the states included for each curve in fig.\ref{Seind}, we first have a look at the running sum of the NME which is a very good tool to understand contributions to NME from different states. As presented in literature\cite{Suhonen:1998ck,Fang:2009ic,Lv:2021oty}, for QRPA, the running sum plays a very important role for our understandings of nuclear structure issues for double beta transition. It is with the SSD or low-lying state dominance (LSD) characteristics for current nucleus, that is the low- and medium-lying states accumulate enough strength much larger than the final strength and it is then got cancelled by high-lying states as shown in fig.\ref{runsum} (Blue and red lines). In this sense, just the low-lying states are already enough to describe the decay strength (running sum). For current calculation, for the running sum of NME, we find cancellation from states around excitation energies of 7 MeV, whose energy is a bit  smaller than the GTR energy, and it comes most probably from a pygmy GTR, and also the cancellation from Giant GTR can be identified in the running sum around excitation energies of 12MeV but their contributions are suppressed by the energy denominator. This is actually well understood for QRPA calculations, so for the next step, we want to study whether the spectra behave like NME, that if there are cancellations presented, the low-lying states alone determine the spectra. Or in other words, could the spectra tell us the information of the cancellation directly. 

To do this, we divide the intermediate states into three parts in fig.\ref{runsum} (We consider the case of QRPA calculations without quenching, the blue serial lines): the low-lying states which already accumulate enough strength equal to the total final strength(the pale blue line), the additive low-lying (medium-lying) states which accumulates more strength reaching the maximum strength  (the light blue line) and the third part is the cancellation to the excess strength from the high-lying states(the blue line). Correspondingly, in fig.\ref{Seind}, we study the contributions of these three parts by comparing the spectra with the pale blue, light blue and blue curves (the counterparts of the three pars infig.\ref{runsum}) respectively. In this way, we try to understand how the additive part and the cancellation part contribute to the spectra. The results are similar to that of NME, while spectra from the low-lying part are close to SSD shape, the additive part will push the spectra away from SSD shape to the HSD shape and the cancellations parts(high-lying states) will pull the spectra back. And the spectra are almost solely determined by the low-lying part which also accumulates enough strength for the running sum in fig.\ref{runsum}(both in pale light curves), this agrees with the conclusion of \cite{Simkovic:2018rdz} where their calculation suggests that the spectra is sensitive to $M^{2\nu}_{GT-3}$ dominated by the low energy contribution. Therefore, if the contribution from medium energy region and high energy region cancels each other (in our case light blue and blue curves), we barely see any implications in the spectra. That means, like the NME, the electron spectra cannot help us distinguish between the true single or low-lying states dominance (no strength from high-lying states) and effective SSD (LSD) (strength from high-lying states get cancelled by each other). 
And here the analysis applies to the normal case, special cases such as the first state contributes more strength than needed or there are flip of the sign of the strength will behave differently. These are coincidently the cases for the other two nuclei we are interested, so we will leave these cases to subsequent sections.  

Now, we proceed to the discussion of electron spectra from QRPA calculations which are presented in fig.\ref{spectraSe}. The spectra from QRPA seems to agree with the measurement, especially for a bare $g_A$. Although the deviations from the head and tail seem to be drastic, they actually contribute less to the counts of the events. In general, current QRPA calculation qualitatively reproduce the experimental results, and a cancellation from high-lying states is expected in current calculation. And future measurement will pin down the errors and give quantitative results as that for $^{136}$Xe (which we will discuss later), these will help us better constrain the QRPA calculations since different calculations are still differentiate by details.

Next, we apply such analysis to the NSM calculations as well. As we may be aware, simultaneous fulfillment of the GT strength of parent and daughter nuclei to the intermediate nuclei and double beta decay strength for QRPA calculations has long been a bothering problem\cite{Faessler:2007hu}. This may be caused by the fact that we mimic the multi-phonon behavior by overestimate the particle-particle correlations, and this over-estimation for different observable are most probably different. This is also why in this study, we don't perform an analysis of the B(GT) strength of the intermediate states to parent and daughter nuclei from QRPA calculations. This will not be the case for NSM calculation since all excitations beyond the one-phonon excitation are included, but NSM may face the problem of missing Giant GTR (or Pygmy one) strength which serves as important source for high energy cancellations to double beta decay NME as predicted from QRPA calculations. Also, for low-lying states, certain quenching is needed to account the missing correlations from outside of the model space.
The usual way of fixing the quenching of $g_A$ for $2\nu\beta\beta$-decay is by fitting the $2\nu\beta\beta$-decay NME  \cite{Brown:2015gsa}. In our current calculation we fit the half-lives instead, and find that $g_A$ is basically the same as that of fitting the NME. Our fitted $g_A\sim 0.55$ is slightly larger than the fitted value for $^{76}$Ge \cite{Brown:2015gsa} with a strength-function method. Also, one observes that the NME or half-life are nearly converge with current chosen number of intermediate states, this agree with various NSM calculations\cite{Li:2017nho,Kostensalo:2022lzk}. However, with such fitting strategy, besides the successful prediction of half-lives, the calculated spectra is not satisfying. In fig.\ref{spectraSe}, both NSM calculations (purple and brown solid lines) favor HSD and contradict current experimental data.

\begin{figure*}
\includegraphics[scale=0.3]{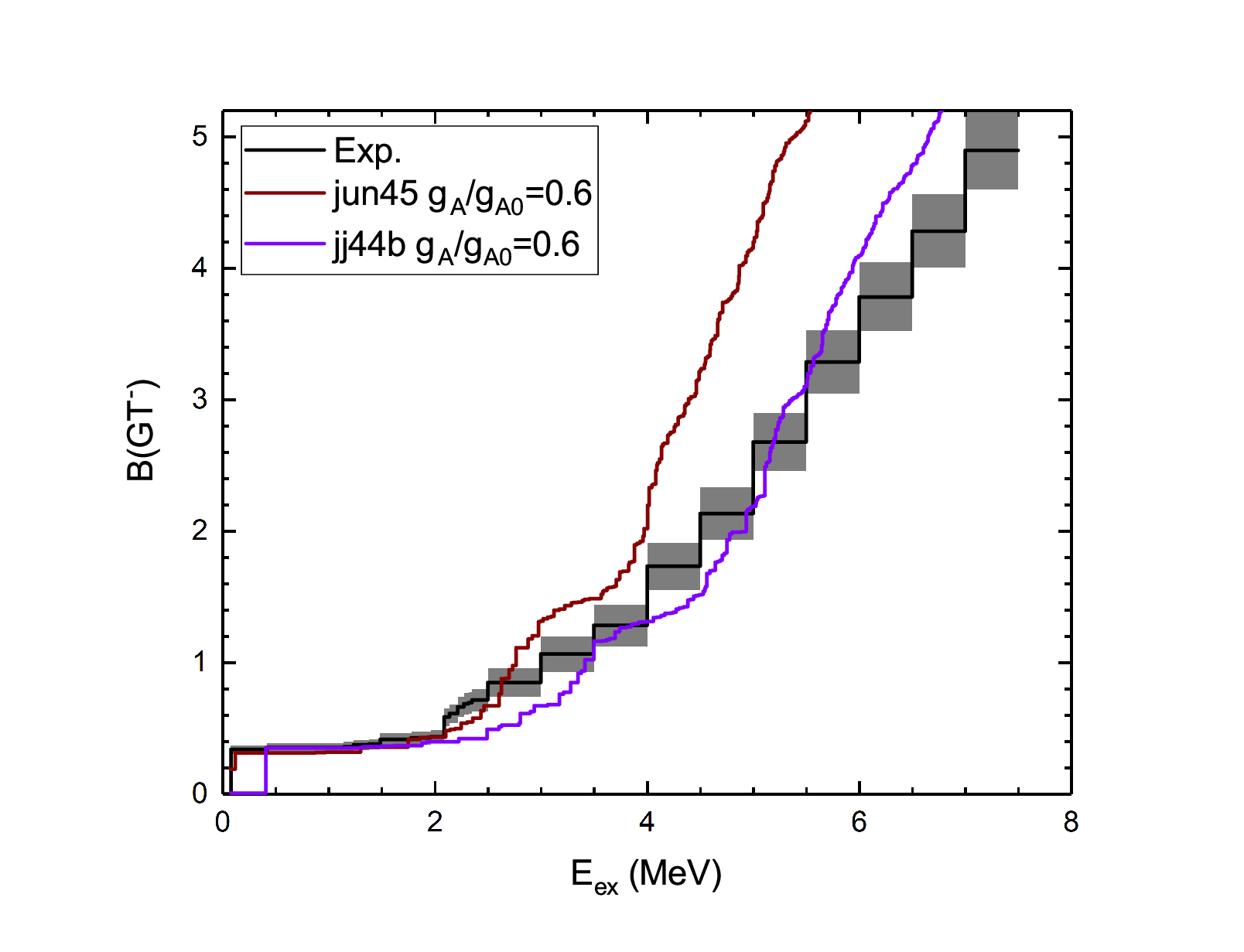}
\includegraphics[scale=0.3]{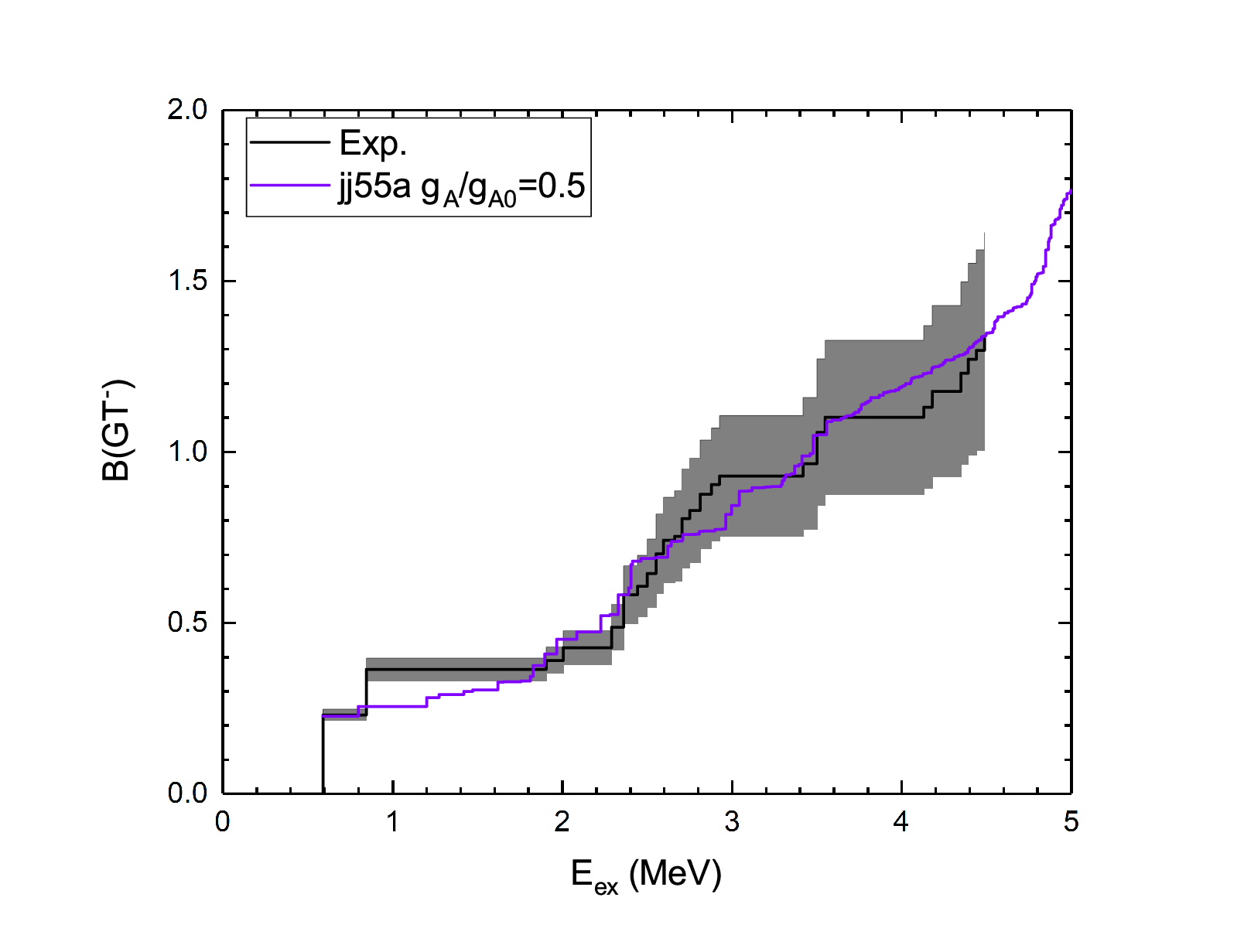}
\caption{(Color online) B(GT$^-$) running sum for $^{82}$Se and $^{136}$Xe respectively. The shadow denotes the errors from charge exchange reaction $^{82}$Se($^{3}$He,t)$^{82}$Br \cite{Frekers:2016haa} or $^{136}$Xe($^{3}$He,t)$^{136}$Cs \cite{Frekers:2018edj}.}
\label{BGT}
\end{figure*}

On the other hand, for $^{82}$Se, the charge exchange experiments $^{82}$Se($^{3}$He,t)$^{82}$Br\cite{Frekers:2016haa} offer relatively precise B(GT) values for the $\beta^-$ side while the data for $\beta^+$ side is still missing (see fig.\ref{BGT}). To reproduce the experiment data of the low-lying strength from NSM calculations, we find a quenched value of $q\equiv g_A/g_{A0}=0.6$ is needed. While for $\mathbf{jun45}$, we find that such a fitting will lead to a large deviation for states with excitation energy larger than 5 MeV. A stronger quenching will make better agreement for these high-lying states, but then the low-lying strength will be heavily suppressed, as we will show, the spectra impose a severe constraints on the decay strength at low excitation energy, and a larger quenching value is therefore needed. But still these fitted $g_A$'s are different from that of double beta decay.

These discrepancies lead to severe problems of reliability of NSM description for double beta decay. A straightfoward solution is that we adopt the same quenching of $g_A$ for both charge-exchange reactions and double beta decay. With the fitted $g_A$ from charge-exchange reactions, from NSM calculations the first few states accumulate enough strength to reproduce the decay strength (light brown and light purple lines in fig.\ref{runsum}).From above analysis, we then need consider these low-lying states only for the spectra. We find these new spectra is getting improved (dashed lines in fig.\ref{spectraSe}), especially for results from $\mathbf{jun45}$ (dashed brown lines). These results also show that, for a SSD-like spectra, the results are extremely sensitive to the very low-lying states, better agreement is achieved by $\mathbf{jun45}$ just because its first two states better reproduce the experimental B(GT) in fig.\ref{BGT}, despite the fact that the description of B(GT) from higher-lying states is not satisfying. In this sense, these kind of spectra can well constrain the strength distribution. $\mathbf{jj44b}$ Hamiltonian comes out to be a bad example as it fails to reproduce the low-lying strength, just several hundred keV deviation of the excitation energy leads to a worse prediction. 
While the spectra implies high-lying cancellation from NSM calculations, we will not get this from NSM calculation even if we perform a full diagonalization of the NSM Hamiltonian. This is due to the lack of the spin-orbit partner of $f_{7/2}$ and $g_{9/2}$. 

In one word, current NSM results with proper treatment validate the simultaneous fulfillment of quenching for both $\beta$ and $\beta\beta$ side and could predict things which are missing in the calculations.  Nevertheless, we still lack $^{82}$Kr charge exchange data before drawing a firm conclusion, and current study can be a good Ansatz for combined analysis on double beta decay and charge exchange reactions. 


Thus, the observed spectra rule out the NSM calculations with $\mathbf{jj44b}$ Hamiltonian although it gives a better agreement for low-lying B(GT). This suggests that the electron spectra can be used for constraining the decay strength of the very low-lying states for the SSD case. The future measurement with a parametrized shape (see discussion in subsequent sections) could help us with a more quantitative analysis.

\subsection{$^{100}$Mo}
\begin{figure*}
\includegraphics[scale=0.5]{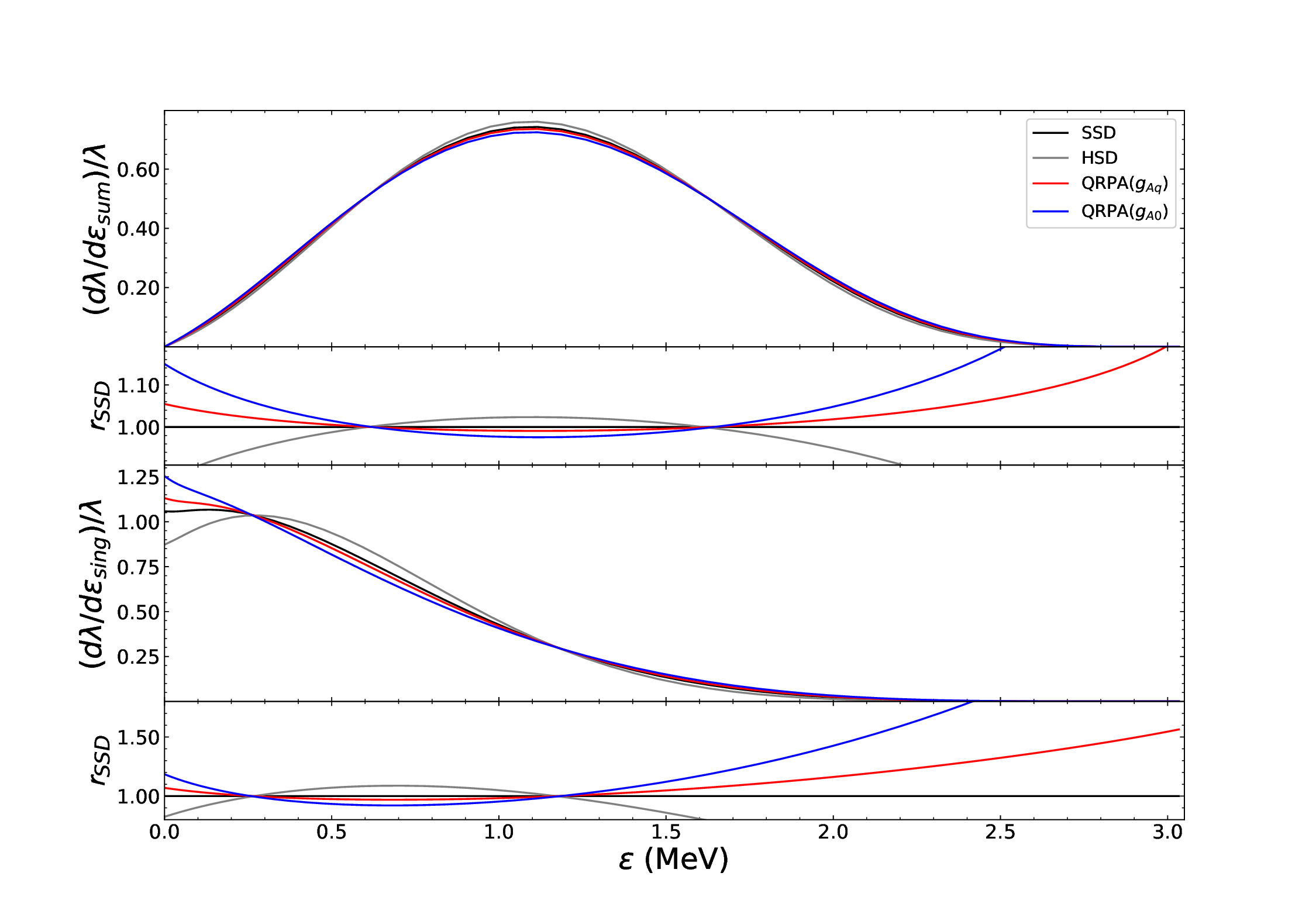}
\caption{(Color online) QRPA calculations of electron spectra for $^{100}$Mo. Here $g_{A0}$ means results without consideration of quenching while $g_{Aq}$ refers to the case of $g_A=0.75g_{A0}$.}
\label{spectraMo}
\end{figure*}

The numerical estimation of electron spectra for $^{100}$Mo has been done by Skyme meanfield based QRPA\cite{Moreno:2008dz}. However, their prediction with HSD trend has now been ruled out by NEMO-3 experiments. Our results differ greatly from theirs with a strong favor going beyond SSD. From fig.\ref{spectraMo}, we find that for fitted $g_{pp}^{T=0}$ with both $g_{A0}$ and $0.75g_{A0}$ cases, the first state contribute more strength than the final decay strength requiring the cancellations from high-lying states. Compared to $^{82}$Se, the cancellations from GTR and other states are somehow weakened, nevertheless we can still find the trace of cancellation from pygmy or giant GTR from fig.\ref{runsum}. As we have mentioned above, it is nearly impossible to simultaneously reproduce the following three quantities in QRPA calculation: B(GT$^-$), B(GT$^{+}$) and $M^{2\nu}_{GT}$\cite{Faessler:2007hu}, but with enlarged $g_{pp}$, we could always mimic the multi-phonon behavior for $M^{2\nu}_{GT}$. Since the ground state of $^{100}$Tc comes out to be $1^+$, we could estimate the $2\nu\beta\beta$-decay strength for the very first excited states from measurements. The analysis in \cite{Moreno:2008dz} suggests, with EC or charge-exchange reaction extracted B(GT$^-$)  and $\beta$-decay extracted B(GT$^+$) of $^{100}$Tc, a NME from the first state smaller than final $2\nu\beta\beta$ NME is given. Nevertheless, later analysis with the improved B(GT$^-$) from EC of $^{100}$Tc \cite{Semenov:2017dgj} suggests that the first $1^+$ state contributes the NME larger than the total $M^{2\nu}_{GT}$\cite{Barabash:2015eza}. This agrees with our current calculations despite  QRPA reproduce a even larger NME value from the first $1^+$ excited state as presented in fig.\ref{runsum} (Here the horizontal short lines at the y axis are the estimated NMEs from the first $1^+$ state from various measurement). Our calculations for NMEs from the first $1^+$ state agree with the analysis in \cite{Semenov:2017dgj} but differs by about 10\% for both cases with or without $g_A$ quenching.

Such behavior of running sum leads to visible effect on electron spectra in fig.\ref{spectraMo}. Unlike for the case of $^{82}$Se, the predicted spectra don't lie in the region between SSD and HSD as one would expect with a traditional PSF+NME treatment (eq.\eqref{rapp}). The calculated spectra go beyond SSD, this means we will have more events at the spectra head or tail and less events around the peak for the summed spectra. This explicitly pointed out the inadequacy of the traditional expression from eq.\eqref{rapp}. While for single electron spectra, they also look differently, more events will be observed at small and large electron energies while less for medium electron energy range. These features has been observed probably by \cite{Azzolini:2019yib} with a simplified calculation (SSD-3 model in \cite{Semenov:2017dgj,Semenov:2018ddv}), and current results actually agree well with the spectra obtained by NEMO experiments, where preference of SSD is observed with $\chi^2$/ndf=1.54 compared to a $\chi^2/$ndf=42.91 from HSD for single electron spectrum. Especially, a delicate analysis shows that a simplified SSD-3 model has a much smaller $\chi^2/$ndf=1.13 than that $\chi^2/$ndf=1.45 of a simple SSD model for the summed electron spectra, despite for single electron spectra, SSD is slightly favored by the experiments. This also confirms the existence of cancellations to the decay strength from states rather than the first one. But still, we are not still not certain about the details of the cancellations.

Therefore, to probe these details, to measure precise spectra may be a way. Also if the B(GT$^{+}$) and B(GT$^-$) strength of the GTRs to parent and daughter nuclei is measured once, we can surely get the idea whether a cancellation from GTR should be presented and what the roles other excited states play. In one word, we will have clear signature from electron spectra for the case beyond SSD. And surely for this case, the cancellation must be happening and by measuring the respective Gamow-Teller strength from GTR, we could get the general idea whether a high-lying state cancelation exists as predicted by most QRPA calculations\cite{Lv:2021oty}. Generally, for $^{100}$Mo, electron spectra can be used to constrain the various QRPA calculations and rule out certain version which fails to give a reasonable running sum of NME.  
 
\subsection{$^{136}$Xe}

\begin{figure*}
\includegraphics[scale=0.5]{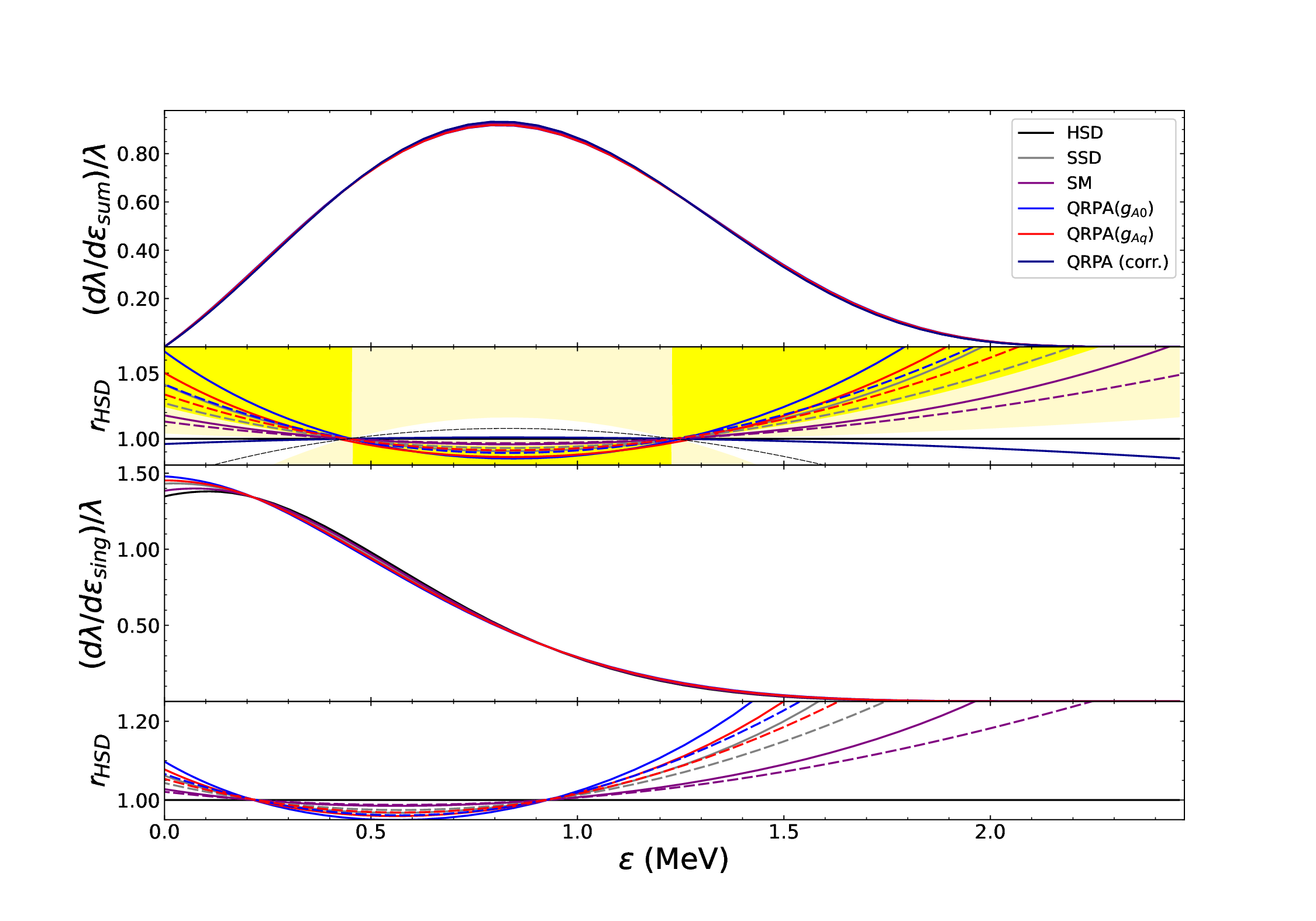}
\caption{(Color online) NSM and QRPA calculations of electron spectra for $^{136}$Xe.  The ratios $r_{HSD}$ here refers to the spectra relative to the HSD shape. Here the legend (QRPA(corr.)) of dark blue line refers to one fit the NME by the negative values illustrated in fig.\ref{runsum}. The black dashed line in second graph is the central value of measured $\xi_{31}^{2\nu}$. The yellow and pale yellow region in the second graph refers to the regions which are excluded at 90\% and 68\% C.L. by KamLand-Zen \cite{KamLAND-Zen:2019imh}. The dashed lines here correspond to the results obtain with TEM, see text.}
\label{spectraXe}
\end{figure*}

The Taylor expansion method (TEM) prescription in \cite{Simkovic:2018rdz} (eq.\eqref{TEM}) actually provide very good parametrization for the electron spectra. As indicated by the phase space factor calculations\cite{Kotila:2012zza}, the spectra, like the phase space factor $G^{2\nu}(\tilde{A})$, converges with $\tilde{A}\ge E_{GTR}$, here $\tilde{A}$ is the energy gap between the first $1^+$ intermediate state and the ground states of parent and daughter nuclei. While, one could easily observe that the 0th order PSF $G^{2\nu}_0$ in \cite{Simkovic:2018rdz} is equivalent to traditional PSF at intermediate energy $\tilde{A}\rightarrow \infty$ or the spectra of HSD. This suggests that the $\xi^{2\nu}_{31}=0$ case for TEM is actually HSD for traditional treatment. And the traditional treatment of PSF covers the parameter space of $\xi^{2\nu}_{31}$ from 0 to a finite positive value corresponding to the positive energy gaps $\tilde{A}$ from infinity to $E_{1}$ defined after eq.\eqref{KL}. As we have seen in $^{100}$Mo, the actual spectra may not be restricted in this parameter space. Current data from KamLAND-Zen tends to give a small positive or even negative $\xi^{2\nu}_{31}$, this contradicts certain many-body approaches. In fig.3 of \cite{KamLAND-Zen:2019imh}, one finds that both methods have been excluded at 1$\sigma$ and part of the QRPA parameter space has been excluded at 2$\sigma$ level.

Our current QRPA calculation agrees with previous ones in \cite{KamLAND-Zen:2019imh}, somehow fails to reproduce the measured spectra. And in this sense, a wrong prediction of cancellations is presented. Actually, in fig.\ref{spectraXe}, difference between the full numerical method (solid lines with corresponding colors) and that of TEM (dashed lines with corresponding colors) is illustrated. We find that TEM has the same trend as full numerical calculations but deviate in detail, the deviation becomes more pronounced when the actual spectra are farer away from HSD case (For current nucleus, unlike the previous two, we compare the spectra with that of HSD because it corresponds to $\xi^{2\nu}_{31}=0$, close to the measured values). The comparison shows that TEM with the missing higher order correction such as $\xi^{2\nu}_{51}$\cite{Simkovic:2018rdz} will give an underestimation of the shift from HSD. In one word, current QRPA calculations with proper quenching factor is hard to reproduce the required spectra as has already been observed.

The situation for NSM is similar, with a very strong $g_A$ quenching ($g_A=0.4$) applied in order to reproduce the experimental half-life, prediction from $\mathbf{jj55a}$ has a strong low-lying strength and the calculated spectra lie in between HSD and SSD, but are somehow excluded by C. L. of  60\%. If we follow the analysis from above sections for $^{82}$Se, as from fig.\ref{BGT},  we get $g_A=0.5 g_{A0}$ by fit the calculated running sum of B(GT) with experiments. In fig.\ref{BGT}, our calculation basically reproduce the first GT strength as well as low-lying strength up to excitation energy 5 MeV. Unfortunately, there is no measurement on the $\beta^+$ side of $^{136}$Ba to further validate our current choice. But this choice of $g_A$ will enhance the low-lying dominance characteristic and worse the situation with the shift of spectra to SSD side.

Above analysis seems to announce the failure of both QRPA and NSM calculations for predicting electron spectra and rule out a strong cancellation most probably from GTR like have been done for previous two nuclei. However, the measurement still leave us space since the central value of measured spectrum shape parameter $\xi^{2\nu}_{31}$ lies at the negative region. Therefore, in the subsequent part, we explore another probability which help us to reproduce the spectra both QRPA and NSM calculations. Of course, this new explanation needs further more precise measurement of the spectra to pin down the errors, it only holds for the case we do have a negative $\xi^{2\nu}_{31}$. 

We start from QRPA calculation, for decades, the fitting of the parameter $g_{pp}^{T=0}$ (early days $g_{pp}$ for both T=0 and T=1) relies on the $2\nu\beta\beta$-decay GT NME. For all these calculation, one begins with a curve starting at $g_{pp}^{T=0}=0$(See for example fig.1 of \cite{Lv:2021oty}). Since there is an arbitrariness in the choice of phase for the NME, one always set the values at $g_{pp}^{T=0}=0$ to be positive and then by default, one chose the $g_{pp}^{T=0}$ value which give a positive NME close to experimental one. The general reason for such strategy could be of the approximate SU(4) symmetry \cite{Simkovic:2018hiq}, which requires a vanishing $M_{GT-cl}^{2\nu}$ and subsequently a positive $M^{2\nu}_{GT}$ and then the negative NME is supposed to be unnatural although it is not collapsed yet for the QRPA equations. However, it is still lack experimental evidence that to what extend such symmetry is exact. If we temporarily loose this restriction, then there actually exists another possibility, that the fixed NME is with different sign as the value at $g_{pp}^{T=0}=0$. When turns to running sum, this means the running sum flips sign when the excitation energy grows up, the high-lying GTR states will drag the strength from positive to negative. Our study suggest there is a small window for $g_{pp}^{T=0}$, that the strength of first state has different sign as the final strength (dark blue lines in fig.\ref{spectraXe}), leading to a negative $\xi^{2\nu}_{31}$ for TEM. Current measurement actually leaves a very narrow window for a positive $\xi_{31}^{2\nu}$ while a large parameter space for negative $\xi_{31}^{2\nu}$ (see the unshaded region in the second panel of fig.\ref{spectraXe}). This implies that a negative NME may be experimentally allowed which could only be described by current treatment or TEM. 

We now present the calculated spectra with a flipped-sign running $M^{2\nu}_GT$ strength distribution in fig.\ref{spectraXe}(dark blue line). We then find that the calculatd spectra goes beyond the HSD pattern, these results agree well with KamLAND-Zen measurement. The spectra lies close to the pattern of obtained central value for measured $\xi^{2\nu}_{31}$ measurement. This implies that for future measurement , once a negative $\xi^{2\nu}_{31}$ is confirmed, it strongly indicates a strong cancellation most probably from GTR flips the sign of the running sum. In this sense, the cancellation confirmed by previous two nuclei exists also for $^{136}$Xe. Then for NSM, current data suggests also high-lying states cancellation to decay strength too. And the discrepancy between the quenched $g_A$ values from $2\nu\beta\beta$-decay strength and charge exchange reaction disappears too. And to reproduce the desired decay strength, we will need a cancellation of the strength around 0.064, measurements of future charge exchange will confirm our conclusion.

Therefore, based on above analysis, future reduction of uncertainty for the spectra measurement combined with charge exchange experiments will definitely give us more hints on the solutions of current discrepancy. In this sense, these spectra can be used to constrain the many-body calculations. If the final measurement of the parameter $\xi_{31}^{2\nu}$ comes out to be positive near 0, then the current QRPA calculation fails to predict the low-lying strength for $2\nu\beta\beta$-decay, so does the NSM calculation, and then these calculations can be ruled out. Meanwhile, if $\xi_{31}^{2\nu}$ comes out to be negative, then current QRPA calculations could reproduce the results, and a large reduction at high-lying states from NSM is our major prediction for this nucleus then.

\section{Conclusion and outlook}
The $2\nu\beta\beta$-decay spectra offer us rich information and could be used to constrain the nuclear structure calculations. Using the numerical treatment, by combining with the charge exchange experiment data, we come to several important conclusions: i) the excitation states of the intermediate nuclei which sum up enough strength to reproduce the experimental decay half-lives determine the spectra, and other high-lying states whose decay strength cancel each other will not contribute to the spectra; ii) this ensures a consistent description of quenching factor for shell model calculations for $^{82}$Se; iii) both shell model and QRPA calculations point to a  reduction from high-lying states possibly from GTRs to the NME for three nuclei concerned. 

The spectra can actually lie beyond the SSD and HSD shapes from the rough treatment, the former has been observed while the latter has never been considered although there are traces from KamLAND-Zen experiment. All our conclusions still need further verification with charge-exchange experiments especially on the $\beta^+$ side of daughter nuclei. And further high precision double beta spectra measurement could help reduce the uncertainty of spectra shape. Together with the charge-exchange experiments, we could test the universality of high-lying state (possibly GTR) cancelations which is common in QRPA calculations.

\section*{acknowledgement}
This work is supported by National Key Research and Development Program of China (2021YFA1601300).

\bibliography{2nspct}

\end{document}